# Estimation of the Birefringence Change in Crystals Induced by Gravitation Field

R.Vlokh and M.Kostyrko

Institute of Physical Optics, 23 Dragomanov St., 79005 Lviv, Ukraine,
e-mail: vlokh@ifo.lviv.ua

## Abstract

The effect of gravitation field of spherically symmetric mass on the birefringent properties of crystals has been analysed. It has been shown that the gravitation field with spherical symmetry can lead to a change of birefringence in anisotropic media.

## Introduction

It has been shown in our previous work (see *R. Vlokh* [1]) that the gravitation field of spherically symmetric mass cannot induce lowering of symmetry of space or polarizable vacuum and appearance of its optical anisotropy. This conclusion has been based on the relations for the refraction index $n$ (or the dielectric impermeability constant $B_{ij} = (1/n^2)_{ij}$) changes under the action of gravitation field of spherically symmetric mass, derived in the weak-field approximation:

$$n = 1 + 2\sqrt{\beta_{ij} M} (g^{1/2}), \qquad (1)$$

$$B_{ij} = 1 - 4\sqrt{\beta_{ij} M} (g^{1/2}). \qquad (2)$$

Here $\beta_{ij} = G/c_0^4$, $c_0$ is the light speed in vacuum and $G$ the gravitation constant. Namely, the square root of the gravitation field strength (the so-called free-fall acceleration), $g^{1/2}$, describes a scalar action which cannot lower symmetry of a medium. Many authors have already considered light propagation in a flat space near a massive body, basing on the idea of distributed dielectric permittivity (or refractive index) of the space treated as a matter. For instance, *R.H. Dicke* has done this on the basis of *Newton* and *Maxwell* equations (see, e.g., [2]), *H.E. Puthoff* [3] has considered the phenomena analysed usually in terms of curved space-time, using the approach of polarizable vacuum, while *K. Nandi* and *A. Islam* [4], *J. Evans* [5] and *Fernando de Felice* [6] have treated optical phenomena in the gravitation field on the basis of "optical-mechanical analogy". Recently *P. Boonserm et al.* [7] have found that the internal stresses in celestial bodies can lead to appearance of the corresponding optical anisotropy and so a necessity for introduction of "effective refractive index tensor". According to this approach, the refraction index can acquire properties of a second-rank tensor, provided that certain conditions are imposed on the gravitation field. It is necessary to emphasize that the refractive index is not a tensorial quantity, unlike the optical-frequency dielectric impermeability constant, which represents a two-rank tensor. It follows from Eqs. (1) and (2) of the study [1] that the light speed depends upon the gravitation field and approaches the $c_0$ value only if the field strength tends to zero. The $G$ quantity represents in fact material (constitutive) coefficients of the flat space (or the corresponding optical medium) and should therefore obey the *Neumann* principle. Being a scalar action, gravitation field of a spherical mass cannot lead to appearance of anisotropy. In case of hypothetical lowering of initially isotropic symmetry of space by the gravitation or the other fields, the coefficient $G$, the *Hubble* constant and the time can get tensorial properties. In frame of this description, the time plays a role of spatial property. Owing to the *Curie* principle, the symmetry group of the flat space should depend on the field configuration and, following the *Neumann* symmetry principle, it should be a subgroup of symmetry group of the time.

Then the following questions appear: if the gravitation field of spherically symmetric mass induces refractive index change for the "free space" or the polarizable vacuum, could this field change refraction indices of the other types of matter, for example, anisotropic crystals? Furthermore, could the optical birefringence of anisotropic media be sensitive to the changes in the gravitation field of spherical symmetry?

## Dependence of birefringence on the gravitation field

At present, measurements of changes in the absolute refractive index values of the order of $10^{-5}$ are a difficult experimental problem. Nonetheless, the methods for experimental determination of the birefringence changes are more sensitive. For instance, a usual compensation method for measuring the birefringence permits one to detect its increment of the order of $10^{-7}$. As we have mentioned above, the gravitation field, as a scalar action, does not induce the optical anisotropy for itself. Thus the question should be made more specific: does it induce any changes of optical anisotropy? Let us follow from Eq. (2) and present the optical indicatrix equation for the crystals of medium symmetry (i.e., those belonging to trigonal, tetragonal and hexagonal symmetry groups) under the perturbation induced by scalar action of the gravitation field. The form of the tensor $\beta_{ij}$ is the same as that of the $B_{ij}$ tensor and it should depend on the point symmetry of the matter. For example, the changes in the birefringence for the crystals belonging to the mentioned medium symmetries would be determined by the quantities displayed in the columns below:

|  | $g^{1/2}$ |
|---|---|
| $\Delta B_1$ | $-4\sqrt{M\beta}$ |
| $\Delta B_2$ | $-4\sqrt{M\beta}$ |
| $\Delta B_3$ | $-4\sqrt{M\beta}$ |
| $\Delta B_4$ | 0 |
| $\Delta B_5$ | 0 |
| $\Delta B_6$ | 0 |

Then the optical indicatrix equation looks as follows:

$$\left(\frac{1}{n_o^2} - 4\sqrt{\beta Mg}\right)x^2 +$$
$$+\left(\frac{1}{n_o^2} - 4\sqrt{\beta Mg}\right)y^2 + \quad (3)$$
$$+\left(\frac{1}{n_e^2} - 4\sqrt{\beta Mg}\right)z^2 = 1$$

One can estimate the birefringence change (e.g., for the quartz crystals) under the action of the gravitation field of Earth measured on the Earth surface ($g$=9.8m/s$^2$):

$$\delta(\Delta n)_{xz} = \delta(\Delta n)_{yz} = 2(n_o^3 - n_e^3)\sqrt{\beta Mg} \ . \ (4)$$

It is easily seen that the gravitation field causes the changes in the refractive indices not only for the so-called "free space" or "polarizable vacuum" but also for, e.g., the ordinary solid-state materials. For the quartz crystals ($\lambda = 627.8 nm$, $n_o = 1.542819$ and $n_e = 1.55128$) in the gravitation field of Earth, one obtains the birefringence increment equal to $\delta(\Delta n)_{xz} = \delta(\Delta n)_{yz} = 8.5 \times 10^{-11}$. Though this value is rather small to be detected with the known optical techniques, the gravitation-induced birefringence changes could become measurable with increasing gravitation fields.

On the other hand, one can see from Eq. (4) that the birefringence increment involves also the initial value of this birefringence. Thus, a suitable choice of single crystals (or some other anisotropic materials) with a large initial birefringence would provide essentially increased induced birefringence, even under the action of the gravitation field on the Earth surface. So, the refractive indices for $TeO_2$ crystals are $n_o = 2.2597$ and $n_e = 2.4119$ at $\lambda = 632.8 nm$ [8]. Then the induced birefringence calculated according to Eq. (4) increases up to the value of $\delta(\Delta n)_{xz} = \delta(\Delta n)_{yz} = 3.5 \times 10^{-9}$. For a comparison, the appropriate birefringences at the surface of Jupiter are equal to $\delta(\Delta n)_{xz} = \delta(\Delta n)_{yz} = 2.38 \times 10^{-9}$ for $SiO_2$ crystals and $\delta(\Delta n)_{xz} = \delta(\Delta n)_{yz} = 0.98 \times 10^{-7}$ for $TeO_2$ crystals. Such the optical birefringence increments can be already measured with the aid of ring lasers (see, e.g., [9]). Indeed, the sensitivity of the latter to nonreciprocal contribution to the refractive index of the sample of length 10cm and ring laser perimeter of 3.4771m is of $\Delta n \sim 10^{-20}$.

## Conclusions

One can see from our analysis that employing of the "effective refractive index" approach for the description of light propagation near the massive bodies unambiguously leads to a necessity of introducing the optical-frequency impermeability and realizing its dependence on the gravitation field strength. Such a description looks quite familiar for the researches in crystal physics. Probably, the most interesting and important conclusion that follows from this approach is that the initial birefringence of anisotropic media can be changed with the action of the gravitation field, even in the case of spherically symmetric mass. Moreover, these changes can be measured with the help of techniques utilizing the ring lasers.